\documentclass [12pt]{article}

\begin{document}
\begin{center}
\textbf{\huge Canonical entropy of three-dimensional BTZ black
hole}
\end{center}

\begin{center}
Zhao Ren\footnote{Corresponding address: Department of Physics,
Shanxi Datong University, Datong 037009, P.R.China; E-mail
address: zhaoren2969@yahoo.com.cn }

Department of Applied Physics, Xi'an Jiaotong University, Xi'an
710049 P.R.China\\ Department of Physics, Shanxi Datong
University, Datong 037009 P.R.China

Zhang Sheng-Li\\
Department of Applied Physics, Xi'an Jiaotong University, Xi'an
710049 P.R.China
\end{center}

\begin{center}
\textbf{\Large Abstract}
\end{center}

Recently, Hawking radiation of the black hole has been studied
using the tunnel effect method. It is found that the radiation
spectrum of the black hole is not a strictly pure thermal
spectrum. How does the departure from pure thermal spectrum affect
the entropy? This is a very interesting problem. In this paper, we
calculate the partition function by energy spectrum obtained from
tunnel effect. Using the partition function, we compute the black
hole entropy and derive the expression of the black hole entropy
after considering the radiation. And we derive the entropy of
charged black hole. In our calculation, we consider not only  the
correction to the black hole entropy due to fluctuation of energy
but also the effect of the change of the black hole charges on
entropy. There is no other hypothesis. Our result is more
reasonable.According to the fact that the black hole entropy is
not divergent, we obtain the lower limit of
Banados-Teitelboim-Zanelli black hole energy. That is, the least
energy of Banados-Teitelboim-Zanelli black hole, which satisfies
the stationary condition in thermodynamics.\\
\textbf{Keywords:} thermal fluctuation, canonical ensemble,
quantum correction, black hole entropy.\\
PACS numbers: 04.70.-s, 04.70.Dy\vspace{0.5cm}\\
\textbf{\large 1. Introduction}

Black holes have an event horizon from which any matter or
information cannot escape. The loss of the information in horizon
region shows that the horizon has a property of entropy. There are
many methods to calculate the entropies of thermodynamic
quantities. And each method means that taking the horizon area of
the black hole as an entropy is self-consistent. It is caused by
their similarity [1,2]. When the horizon area of the black hole
has been taken as entropy, the energy and temperature of the black
hole satisfy thermodynamic law. Since Hawking radiation has been
discovered, this similarity is described quantificationally [3].
But, how to measures the microstate of the black hole by entropy?
This problem has not been solved. At present, discussing the
entropy of black hole-matter coupling system becomes a meaningful
problem. This problem may provide a way for solving the difficulty
of quantum gravitation. Many researchers have expressed a vested
interest in fixing the relation between statistical mechanic
entropy and thermodynamic entropy [4-8].

In recent years, string theory and Loop quantum gravity both had
succeeded statistically in explaining the black hole entropy-area
law [9-10]. However, which one is perfect? It is expected to
choose it by discussing the quantum correction term of the black
hole entropy. Therefore, studying the black hole entropy
correction value becomes the focus of attention. Many ways of
discussing the black hole entropy correction value have emerged$^{
}$[11-18]. Based on string theory and Loop quantum gravity, the
relationship of the black hole entropy-area is given by [19].

\begin{equation}
\label{eq1} S = \frac{A}{4L_p^2 } + \rho \ln \frac{A}{4L_p^2 } +
O\left( {\frac{L_p^2 }{A}} \right),
\end{equation}

\noindent where $A = 16\pi L_p^2 M^2$ is the area of the black
hole horizon, $L_p = \sqrt {\hbar G} $ is Planck length .Refs.[13]
obtained $ \rho = - 1 / 2 $ in four dimension space-time.

On the other hand, there has been much attention devoted to the
lower-dimensional gravitation theory region. Recently, the study
of two-dimensional black hole thermodynamics shows that entropy
satisfies the area relation and the second law of thermodynamics
[20-22]. In 1992 Banados, Teitelboim and Zanelli (BTZ)
[23,24]showed that (2+1)-dimensional gravity has a black hole
solution. This black hole is described by two (gravitational)
parameters, the mass $M$ and the angular momentum $J$. It is
locally AdS and thus it differs from the Schwarzschild and Kerr
solutions since it is an asymptotically anti-de Sitter instead of
a flat space-time. Additionally, it has no curvature singularity
at the origin. AdS black holes are members of this two-parametric
family of BTZ black holes, and they are very interesting in the
framework of string theory and black hole physics [25,26].

Recently, a new explanation for Hawking radiation process of the
black hole is tunnel process. Based on it, the radiation spectrum
is derived. However, this radiation spectrum has a departure from
pure thermal spectrum. How does it affect the black hole entropy?
In this paper, we calculate the partition function of BTZ black
hole using the radiation spectrum obtained in tunnel process.
Furthermore, we derive the entropy of canonical black hole. In our
calculation, there is no the other hypothesis. We provide a new
way for discussing the entropy of canonical black hole.We take the
simple
function form of temperature ($c = K_B = 1)$.\vspace{0.5cm}\\
\textbf{\large 2. Canonical partition function }

Parikh and Wilczek [27] discussed Hawking radiation by tunnel
effect. They thought that tunnels in the process of the particle
radiation had no potential barrier before particles radiated.
Potential barrier is produced by radiation particles itself. That
is, during the process of tunnel effect creation, the energy of
the black hole decreases and the radius of the black hole horizon
reduces. The horizon radius becomes a new value that is smaller
than the original value. The decrease of radius is determined by
the value of energy of radiation particles. There is a classical
forbidden band-- potential barrier between original radius and the
one after the black hole radiates. Parikh and Wilczek skillfully
obtained the radiation spectrum of Schwarzschild and
Reissner-Nordstrom black holes. Refs.[28-37] developed the method
proposed by Parikh and Wilczek. They derived the radiation
spectrum of the black hole in all kinds of space-time.
Refs.[38-41] obtained radiation spectrum of Hawking radiation
after considering the generalized uncertainty relation. And
Angheben, Nadalini, Vanzo and Zerbini have computed the radiation
spectrum of the arbitrary dimensional black hole and obtained the
energy spectrum of radiation particles of general black hole
[37,42]

\begin{equation}
\label{eq2} \rho _s \propto e^{\Delta S},
\end{equation}

\noindent where

\[
\Delta S = S_{MC} (E - E_s ) - S_{MC} (E)
 = \sum\limits_{k = 1} {\frac{1}{k!}\left( {\frac{\partial ^kS_{MC}
(E_b )}{\partial E_b^k }} \right)_{E_s = 0} } ( - E_s )^k
\]

\begin{equation}
\label{eq3}
 = - \beta E_s + \beta _2 E_s^2 + \cdots ,
\end{equation}
where, $E_b = E - E_s $, $E_s $ is energy of radiated particle,
 $S_{MC} (E - E_s )$ is entropy of Microcanonical
ensemble with energy $(E - E_s )$ (namely Bekenstein-Hawking
entropy of black hole with energy $(E - E_s )$). According to the
relation of thermodynamics, $\beta $ should be the inverse of the
temperature.

\begin{equation}
\label{eq4} \beta _k = \frac{1}{k!}\left( {\frac{\partial ^k\ln
\Omega }{\partial E_b }} \right)_{E_s = 0}
 = \frac{1}{k!}\left( {\frac{\partial ^kS_{MC} }{\partial E_b }}
\right)_{E_s = 0} .
\end{equation}
Normalizing the distribution function $\rho _s $, we obtain $\rho
_s = \frac{1}{Z_c }e^{\Delta S}$, where

\begin{equation}
\label{eq5} Z_c = \sum\limits_{E_s}\rho (E - E_s )e^{S_{MC} (E -
E_s ) - S_{MC} (E)}
\end{equation}

\noindent is the partition function. For the semi-classical
thermal equilibrium system, the canonical partition function can
be expressed as

\begin{equation}
\label{eq6} Z_c (\beta ) = \int\limits_0^\infty {e^{\Delta S}dE_s
} \rho (E - E_s ),
\end{equation}

\noindent where $\rho (E - E_s )$ is a density of state of
ensemble (black hole) with energy $(E - E_s )$ . From Ref.[43], we
have $\rho (E - E_s ) \equiv e^{S_{MC} (E - E_s )}$.

The canonical entropy is expanded a Taylor series near energy $E$,

\begin{equation}
\label{eq7} S_{MC} (E - E_s ) = S_{MC} (E) - \beta E_s + \beta _2
E_s^2 + \cdots .
\end{equation}
When we neglect the higher-order small term, Eq.(\ref{eq6}) can be
rewritten as

\[
Z_c (\beta ) = \int\limits_0^\infty {e^{ - \beta E_s + \beta _2
E_s^2 }dE_s } e^{S_{MC} (E - E_s )}
 = e^{S_{_{MC} } (E)}\int\limits_0^\infty dE_s{e^{ - 2\beta E_s +
2\beta _2 E_s^2 }}
\]

\begin{equation}
\label{eq8}
 = e^{S_{MC} (E)}
\left[ {\frac{1}{2}\sqrt {\frac{\pi }{ - 2\beta _2 }} \exp \left(
{\frac{\beta ^2}{ - 2\beta _2 }} \right)\left( {1 - erf\left(
{\frac{\beta }{\sqrt { - 2\beta _2 } }} \right)} \right)} \right].
\end{equation}

\noindent where

\[
erf(z) = \frac{2}{\sqrt \pi }\int\limits_0^z {e^{ - t^2}} dt
\]

\noindent
is error integral.\vspace{0.5cm}\\
\textbf{\large 3. Canonical entropy}

According to the relation between the partition function and
entropy

\begin{equation}
\label{eq9} S = \ln Z - \beta \frac{\partial \ln Z}{\partial \beta
},
\end{equation}

\noindent we obtain that the entropy of the canonical system is

\begin{equation}
\label{eq10} S_C (E) = S_{MC} (E) + \Delta _S ,
\end{equation}

\noindent where

\begin{equation}
\label{eq11} \Delta _S = \ln f(\beta ) - \beta \frac{\partial \ln
f(\beta )}{\partial \beta },
\end{equation}

\begin{equation}
\label{eq12} f(\beta ) = \frac{1}{2}\sqrt {\frac{\pi }{ - 2\beta
_2 }} \exp \left( {\frac{\beta ^2}{ - 2\beta _2 }} \right) \left[
{1 - erf\left( {\frac{\beta }{\sqrt { - 2\beta _2 } }} \right)}
\right].
\end{equation}
According to the asymptotic expression of the error function

\[
erf(z) = 1 - \frac{e^{ - z^2}}{\sqrt \pi z}\left[ {1 +
\sum\limits_{k = 1}^\infty {( - 1)^k\frac{(2k - 1)!!}{(2z^2)^k}} }
\right], \quad \left| z \right| \to \infty ,
\]

\noindent we have

\begin{equation}
\label{eq13} f(\beta ) = \frac{1}{2\beta }\left[ {1 +
\sum\limits_{k = 1}^\infty {( - 1)^k\frac{(2k - 1)!!}{2^k}\left(
{\frac{\sqrt { - 2\beta _2 } }{\beta }} \right)} ^{2k}} \right].
\end{equation}
Substituting (\ref{eq13}) into (\ref{eq11}), we derive

\[
\Delta _S = \ln \left[ {\frac{1}{2\beta } + \sum\limits_{k =
1}^\infty {( - 1)^k\frac{(2k - 1)!!}{2^k2\beta }\left(
{\frac{\sqrt { - 2\beta _2 } }{\beta }} \right)^{2k}} } \right]
\]

\begin{equation}
\label{eq14}
 + \frac{1 + \sum\limits_{k = 1}^\infty {( - 1)^k(2\sqrt { - 2\beta _2
} )^{2k}\frac{(2k + 1)(2k - 1)!!}{2^k(2\beta )^{2k}}} }{1 +
\sum\limits_{k = 1}^\infty {( - 1)^k(2\sqrt { - 2\beta _2 }
)^{2k}\frac{(2k - 1)!!}{2^k(2\beta )^{2k}}} }.
\end{equation}
The thermal capacity of the system is

\begin{equation}
\label{eq15} C \equiv - \beta ^2\left( {\frac{\partial E}{\partial
\beta }} \right),
\end{equation}

\noindent and

\begin{equation}
\label{eq16} \beta _2 = - \frac{1}{2}\frac{\beta ^2}{C}.
\end{equation}
Then Eq.(\ref{eq14}) can be expressed as

\begin{equation}
\label{eq17} \Delta _S = \ln \left[ {T + T\sum\limits_{k =
1}^\infty {( - 1)^k\frac{(2k - 1)!!}{2^kC^k}} } \right]
 + \frac{1 + \sum\limits_{k = 1}^\infty {( - 1)^k\frac{(2k + 1)(2k -
1)!!}{2^kC^k}} }{1 + \sum\limits_{k = 1}^\infty {( - 1)^k\frac{(2k
- 1)!!}{2^kC^k}} }.
\end{equation}
where $T$ is the temperature of the system. When we only consider
the logarithmic correction term

\begin{equation}
\label{eq18} \Delta _S \approx \ln \left[ {T + T\sum\limits_{k =
1}^\infty {( - 1)^k\frac{(2k - 1)!!}{2^kC^k}} } \right].
\end{equation}
In error function, we take the sum $k$ from one to $n$ as the
approximate value of the series. When $z$ is a real number, its
error does not exceed the absolute value of the first term
neglected in the series. Therefore, when $C < - 1$ or $C > 1$, the
first term in $\Delta _S $ is not divergent.\vspace{0.5cm}\\
\textbf{\large 4. Canonical entropy of BTZ black hole}

For the non-rotating Banados-Teitelboim-Zanelli (BTZ) black hole
[23]

\begin{equation}
\label{eq19} ds^2 = - \left( { - M + \frac{r^2}{l^2} +
\frac{J^2}{4r^2}} \right)dt^2 + \left( { - M + \frac{r^2}{l^2} +
\frac{J^2}{4r^2}} \right)^{ - 1}dr^2 + r^2\left( {d\theta -
\frac{J}{2r^2}dt} \right)^2,
\end{equation}

\noindent where, $M$ is the Arnoeitt-Deser-Misner (ADM) mass, $J$
is the angular momentum (spin) of the BTZ black hole, $l^2 = 1 /
\Lambda ^2$ and $\Lambda $ is the cosmological constant.

The outer and inner horizon, i.e., $r_ + $ (henceforth simply the
black hole horizon) and $r_ - $ respectively, concerning the
positive mass black hole spectrum with spin ($J \ne 0)$ of the
line element (\ref{eq19}) are given by

\begin{equation}
\label{eq20} r_\pm ^2 = \frac{l^2}{2}\left( {M\pm \sqrt {M^2 -
\frac{J^2}{l^2}} } \right),
\end{equation}

\noindent and therefore, in terms of the inner and outer horizons,
the black hole mass and the angular momentum are given,
respectively, by

\begin{equation}
\label{eq21} M = \frac{r_ + ^2 }{l^2} + \frac{J^2}{4r_ + ^2 },
\quad J = \frac{2r_ + r_ - }{l}.
\end{equation}
The Hawking temperature $T_H $ of the black hole horizon is [44]

\begin{equation}
\label{eq22} T_H = \frac{1}{2\pi r_ + }\sqrt {\left( {\frac{r_ +
^2 }{l^2} + \frac{J^2}{4r_ + ^2 }} \right)^2 - \frac{J^2}{l^2}}
 = \frac{1}{2\pi r_ + }\left( {\frac{r_ + ^2 }{l^2} - \frac{J^2}{4r_ +
^2 }} \right).
\end{equation}
In two space-time dimensions we do not have an area law for the
black hole entropy; however, one can use a thermodynamic reasoning
to define the entropy [44]

\begin{equation}
\label{eq23} S_{MC} = 4\pi r_ + .
\end{equation}
The specific heat of the black hole is given by [45]

\begin{equation}
\label{eq24} C = \frac{dE}{dT_H } = \frac{dM}{dT_H } = 4\pi r_ +
\left( {\frac{r_ + ^2 - r_ - ^2 }{r_ + ^2 + 3r_ - ^2 }} \right) =
S_{MC} \left( {\frac{r_ + ^2 - r_ - ^2 }{r_ + ^2 + 3r_ - ^2 }}
\right).
\end{equation}
When $r_ + > > r_ - $, $C \approx S_{MC} $, based on the
Eq.(\ref{eq22}) $T_H \approx \frac{S_{MC}}{8\pi^2 l^2}$, then
Eq.(\ref{eq18}) can be rewritten as

\begin{equation}
\label{eq25} \Delta _S \approx  \ln S_{MC} + \ln \left[ {1 +
\sum\limits_{k = 1} {( - 1)^k\frac{(2k - 1)!!}{2^kS_{MC}^k }} }
\right] + const.
\end{equation}
Based on the Eq.(\ref{eq25}), when $S_{MC} > 1$, the entropy is
not divergent. Therefore, we can obtain that the energy of BTZ
black hole satisfies the following condition.

\begin{equation}
\label{eq26} r_ + > \frac{1}{4\pi }.
\end{equation}
From Eq.(\ref{eq21}), when $r_ + > > r_ - $, $r_ + ^2 \approx
Ml^2$. So when the black hole energy satisfies $M > \left(
{\frac{1}{4\pi l}} \right)^2$, the entropy is not divergent.
Otherwise, the entropy is divergent. It means that the black hole
is not at the thermodynamic stationary state. That is, in universe
there is no BTZ black hole with energy that is smaller than this
energy.

According to Eq.(\ref{eq24}), when $r_ + \to r_ - $, $C \to 0$,
the logarithmic term in entropy is divergent. So we obtain that
the black hole is mechanic unstable in this case. BTZ black hole
cannot become an extreme black hole by adjusting the value of $J$.
The extreme black hole exists at the very start of the universe.
This is consistent with the view of
Refs.[46,47].\vspace{0.5cm}\\
\textbf{\large 5. Conclusion and discussion }

Ref.[11] obtained the following result, when they discussed the
correction to entropy of Schwarzschild black hole by the
generalized uncertainty relation.

\begin{equation}
\label{eq27} S
 = \frac{A}{4} - \frac{\pi \alpha ^2}{4}\ln \left( {\frac{A}{4}}
\right) + \sum\limits_{n = 1}^\infty {C_n } \left( {\frac{A}{4}}
\right)^{ - n} + const.
\end{equation}
According to Eq.(\ref{eq27}), there is a uncertain factor $\alpha
^2$ in the logarithmic term in the correction to entropy. However,
in our result, there is no uncertain factor in Eq.(\ref{eq25}).

After considering the correction to the black hole thermodynamic
quantities due to thermal fluctuation, the expression of entropy
is [48-50]

\begin{equation}
\label{eq28} S = \ln \rho = S_{MC} - \frac{1}{2}\ln (CT^2) +
\cdots ,
\end{equation}
There is a limitation in the above result. That is the thermal
capacity of Schwarzschild black hole is negative. This leads to
the logarithmic correction term divergent given by
Eq.(\ref{eq28}). So this relation is not valid to Schwarzschild
black hole. However, for general four-dimensional curved
space-times, when we take a proper approximation or limit, they
can return to Schwarzschild space-times. This implies that
Eq.(\ref{eq28}) has not universality. However, in our result we
only request the thermal capacity satisfies $C < - 1$ or $C > 1$.
According to the discussion to Schwarzschild black hole, we obtain
that this condition may be the condition that the black hole
exists.

In addition, the research of the black hole entropy is based on
the fact that the black hole has thermal radiation and the
radiation spectrum is a pure thermal spectrum. However, Hawking
obtained that the radiation spectrum is a pure thermal spectrum
only under the condition that the background of space-time is
invariable. During this radiation process, there exist the
information loss. The information loss of the black hole means
that the pure quantum state will disintegrate to a mixed state.
This violates the unitarity principle in quantum mechanics. When
we discuss the black hole radiation by the tunnel effect method,
after considering the conversation of energy and the change of the
horizon, we derive that the radiation spectrum is no longer a
strict pure thermal spectrum. This method can avoid the limit of
Hawking radiation and point out that the self-gravitation provides
the potential barrier of quantum tunnel.

Our discussion is based on the quantum tunnel effect of the black
hole radiation. So our discussion is very reasonable. We provide a
way for studying the quantum correction to Bekenstein-Hawking
entropy. Based on our method, we can further check the string
theory and single Loop quantum gravity and determine which one is
perfect. When the thermal capacity of the black hole satisfies $0
\le C \le 1$, the logarithmic correction term of the black hole
may be divergent. For general black hole, it needs further discuss
that this divergent implies physics
characteristic.\\
ACKNOWLEDGMENT

This project was supported by the National Natural Science
Foundation of China under Grant No. 10374075 and the Shanxi
Natural Science Foundation of China under Grant No.
2006011012;20021008.

\textbf{REFERENCES}

[1] J. D. Bekenstein, \textit{Phys. Rev}. \textbf{D7}, 2333 (1973)

[2] J. D. Bekenstein, \textit{Phys. Rev}. \textbf{D9}, 3292 (1974)

[3] S. W. Hawking, \textit{Commun Math Phys}. \textbf{43}, 199
(1975)

[4] L. Susskind and J. Uglum, \textit{Phys. Rev.} \textbf{D50},
2700 (1994)

[5] J. G. Demers, R. Lafrance and R. C. Myers, \textit{Phys. Rev}.
\textbf{D52}, 2245(1995)

[6] `t Hooft G Nucl. Phys. \textbf{B256}, 727 (1985)

[7] M. Kenmoku, K. Ishimoto and K. K. Nandi, \textit{Phys. Rev}.
\textbf{D73}, 064004 (2006)

[8] L. C. Zhang and R. Zhao, \textit{Acta Phys. Sin.}\textbf{ 53},
362(2004) (in Chinese)

[9] A. Strominger and C. Vafa, \textit{Phys. Lett.} \textbf{B379},
99(1996)

[10] A. Ashtekar, J. Baez, A. Corichi, and K. Krasnov, Phys. Rev.
Lett. \textbf{80}, 904(1998)

[11] A. J. M. Medved and E. C .Vagenas, \textit{Phys.
Rev.}\textbf{D70}, 124021(2004)

[12] A. Chatterjee and P. Majumdar, \textit{Phys. Rev. Lett.}
\textbf{92}, 141301(2004)

[13] A. Ghosh and P. Mitra, \textit{Phys. Rev}.\textbf{D71},
027502(2005)

[14] S. Mukherji and S. S. Pal, \textbf{JHEP0205}, 026(2002)

[15] A. Chatterjee and P. Majumdar, \textit{Phys. Rev}.
\textbf{D71}, 024003 (2005)

[16] Y. S. Myung, Phys. Lett. \textbf{B579}, 205(2004)

[17] M. M. Akbar and S. Das, \textit{Class. Quant. Grav}.
\textbf{21}, 1383 (2004)

[18] S. Das, \textit{Class. Quant. Grav}. \textbf{19}, 2355 (2002)

[19] G. A. Camelia, M. Arzano and A. Procaccini, \textit{Phys.
Rev}. \textbf{D70}, 107501(2004)

[20] R. C. Myers, \textit{Phys. Rev}. \textbf{D50}, 6412 (1994)

[21] J. G. Russo, \textit{Phys. Lett}. \textbf{B359}, 69 (1995)

[22] J. D. Hayward, \textit{Phys. Rev}. \textbf{D52}, 2239 (1995)

[23] M. Banados, C. Teitelboim and J Zanelli, \textit{Phys. Rev.
Lett}. \textbf{69}, 1849(1992)

[24] M. Banados, M. Henneaux, C. Teitelboim and J. Zanelli,
\textit{Phys. Rev}. \textbf{D48}, 1506(1993)

[25] J. Maldacena, J. Michelson and A. Strominger, \textbf{JHEP
9902}, 011(1999)

[26] M. Spradlin and A. Strominger, \textbf{JHEP9911}, 021(1999)

[27] M. K. Parikh and F. Wilczek, \textit{Phys. Rev. Lett}.
\textbf{85}, 5042(2000)

[28] E. C. Vagenas, \textit{Phys. Lett}. \textbf{B503}, 399(2001)

[29] E. C. Vagenas, \textit{Mod. Phys. Lett}. \textbf{A17},
609(2002)

[30] E. C. Vagenas, \textit{Phys. Lett.} \textbf{B533}, 302(2002)

[31] A. J. Medved, \textit{Class. Quant. Grav}. \textbf{19,
}589(2002)

[32] M. K. Parikh, \textit{Phys. Lett}. \textbf{B546}, 189(2002)

[33] A. J. Medved,\textit{ Phys. Rev}. \textbf{D66}, 124009(2002)

[34] E. C. Vagenas, \textit{Phys. Lett}. \textbf{B559}, 65(2003)

[35] J. Y. Zhang and Z. Zhao, \textit{Phys. Lett}. \textbf{B618},
14(2005)

[36] J. Y. Zhang and Z. Zhao, \textit{Nucl. Phys}. \textbf{B725},
173 (2005);\textbf{ JHEP10},\textbf{ }055(2005)

[37] M. Angheben, M. Nadalini, L. Vanzo and S. Zerbini,
\textbf{JHEP05}, 014(2005)

[38] M. Arzan, A. J. M. Medved and E. C. Vagenas, \textbf{JHEP09},
037(2005)

[39] A. J. M. Medved and E. C. Vagenas, \textit{Mod. Phys.
Lett}.\textbf{A20}, 2449(2005)

[40] A. J. M. Medved and E. C. Vagenas, \textit{Mod. Phys. Lett.}
\textbf{A20}, 1723(2005)

[41] M. Arzano, \textit{Mod. Phys. Lett.} \textbf{A21},41(2006)

[42] R. Zhao, L. C. Zhang and S. Q. Hu,\textit{ Acta. Phys. Sin.}
\textbf{55}, 3898(2006) (in Chinese)

[43] A. Chatterjee and P. Majumdar gr-qc/0303030

[44] A. Kumar and K. Ray, \textit{Phys. Lett.} \textbf{B351}, 431
(1995)

[45] M. R. Setare, \textit{Eur. Phys}. J. C \textbf{33},555 (2004)

[46] S. W. Hawking, G. T. Horowitz and S. F. Ross, \textit{Phys.
Rev}. \textbf{D51}, 4302(1995)

[47] C. Teitelboim, \textit{Phys. Rev.} \textbf{D51}, 4315(1995)

[48] M. R. Setare, \textit{Phys. Lett}. \textbf{B573}, 173(2003)

[49] M. Cavaglia and A. Fabbri, \textit{Phys. Rev}. \textbf{D65},
044012(2002)

[50] G. Gour and A. J. M. Medved, \textit{Class. Quant. Grav.}
\textbf{20}, 3307(2003)

\end{document}